\documentclass[10pt,twocolumn,twoside]{article}
\usepackage[utf8]{inputenc}
\usepackage{todonotes}
\usepackage{color,soul}
\usepackage{subcaption}
\usepackage{amssymb}
\usepackage{enumitem}\setlist{nosep}
\usepackage[section]{placeins}

\begin{document}
\title{U-PASS: an Uncertainty-guided deep learning Pipeline for Automated Sleep Staging}
\author{Elisabeth R. M. Heremans, Nabeel Seedat, Bertien Buyse,\\Dries Testelmans,
Mihaela van der Schaar,
Maarten De Vos 
\thanks{This research is funded by a PhD fellowship from the Research Foundation
- Flanders (FWO) for E. Heremans (FWO project number 1SC2921N) and by the Flemish Government (AI
Research Program).}
\thanks{E. R. M. Heremans and M. De Vos are with KU Leuven, Department of
Electrical Engineering (ESAT), STADIUS Center for Dynamical Systems, Signal Processing and Data Analytics, Kasteelpark Arenberg 10,
B-3001 Leuven, Belgium (e-mail: elisabeth.heremans@kuleuven.be,
maarten.devos@kuleuven.be).}
\thanks{N. Seedat and M. van der Schaar are with the University of Cambridge, Cambridge CB2 1TN, U.K.}
\thanks{B. Buyse and D. Testelmans are with UZ Leuven, Department of Pneumology, Herestraat 49, B-3000 Leuven, Belgium}
}
\date{May 2023}

\maketitle

\emph{Abstract -
    As machine learning becomes increasingly prevalent in critical fields such as healthcare, ensuring the safety and reliability of machine learning systems becomes paramount. A key component of reliability is the ability to estimate uncertainty, which enables the identification of areas of high and low confidence and helps to minimize the risk of error. In this study, we propose a machine learning pipeline called U-PASS tailored for clinical applications that incorporates uncertainty estimation at every stage of the process, including data acquisition, training, and model deployment. The training process is divided into a supervised pre-training step and a semi-supervised finetuning step. We apply our uncertainty-guided deep learning pipeline to the challenging problem of sleep staging and demonstrate that it systematically improves performance at every stage. By optimizing the training dataset, actively seeking informative samples, and deferring the most uncertain samples to an expert, we achieve an expert-level accuracy of 85\% on a challenging clinical dataset of elderly sleep apnea patients, representing a significant improvement over the baseline accuracy of 75\%. U-PASS represents a promising approach to incorporating uncertainty estimation into machine learning pipelines, thereby improving their reliability and unlocking their potential in clinical settings.}\\
    

\section{Introduction}
Machine learning has ushered in a new era in healthcare, providing opportunities for remote monitoring and computer-aided diagnosis. Although machine learning models are performing with human-expert-level accuracy in controlled tasks \cite{Yang2021, Egger2022, Phan2018a, Phan2020, Guillot2021a, Perslev2021}, machine learning models have seen limited integration into clinical practice. The challenges to embedding these models in healthcare have shifted to providing guarantees of reliability, explainability and estimates of uncertainty \cite{Begoli2019, Kundu2021}. These factors are essential to build trust with clinicians and promote adoption of machine learning in clinical settings \cite{9153891, Phan2021, miotto2017}. Uncertainty or confidence estimates 
are needed at all stages of the machine learning workflow: in data acquisition, at training time, and at deployment time. 
In data acquisition and in training, it is important to evaluate the usefulness of the information that is utilized to train the machine learning model. Feeding too much data in a machine learning model is sub-optimal both financially and in terms of model performance \cite{yang2023dataset,MontesdeOcaZapiain2022}. Medical data is often expensive to acquire, requiring specialized equipment to measure each parameter, and expert knowledge to interpret the data. Moreover, feeding uninformative data to the model can deteriorate the performance because it adds noise without adding information \cite{Kwon2022}. The same principles apply to active learning \cite{settles2009active, Nguyen2022}, a training strategy in which the model actively queries an expert to label extra data points which would be most informative to improve the model. 
At deployment time, typical deep learning models always make a prediction, no matter how uncertain they are. Such systems are not usable in high-stakes medical applications: if the machine learning model doesn't indicate when a prediction is uncertain, a clinician cannot rely on it, as all outputs are equally likely to be incorrect \cite{Kilian2021}.
Consequently, we propose that uncertainty estimation provides a principled approach to solving the aforementioned challenges 
in order to successfully integrate a machine learning workflow in the clinic.\\

A variety of uncertainty estimation methods have been studied in the context of machine learning and deep learning \cite{Abdar2020}. In this work, we 
take a leap forward, leveraging the uncertainty estimates to maximize performance and reliability of machine learning models. We aim to develop a framework that improves predictions at every stage of the machine learning process. It should be model-agnostic, as it needs to be applicable to various clinical machine learning models and problems. We fulfill these goals by building an uncertainty-guided machine learning pipeline. Our pipeline follows a data-centric approach, and integrates uncertainty into  data acquisition, training, active learning, and model deployment. 
We show how this greatly improves machine learning predictions and overall model reliability at different levels: 
\begin{enumerate}
  \setlength{\itemsep}{0pt}
  \setlength{\parskip}{0pt}
  \setlength{\parsep}{0pt}
\item we improve the model through the quality of the training data, 
\item we finetune the model to make better predictions on difficult examples, and 
\item we reduce mistakes by deferring predictions on ambiguous data.
\end{enumerate} Our pipeline represents a crucial step towards the integration of machine learning in clinical settings. We demonstrate its effectiveness on the challenging task of sleep staging, a clinical annotation task characterized by high levels of uncertainty, and known for its relatively low inter-rater agreement \cite{Phan2021, VanGorp2022}. \\

\begin{figure*}[]
    \centering
    \includegraphics[width=\textwidth]{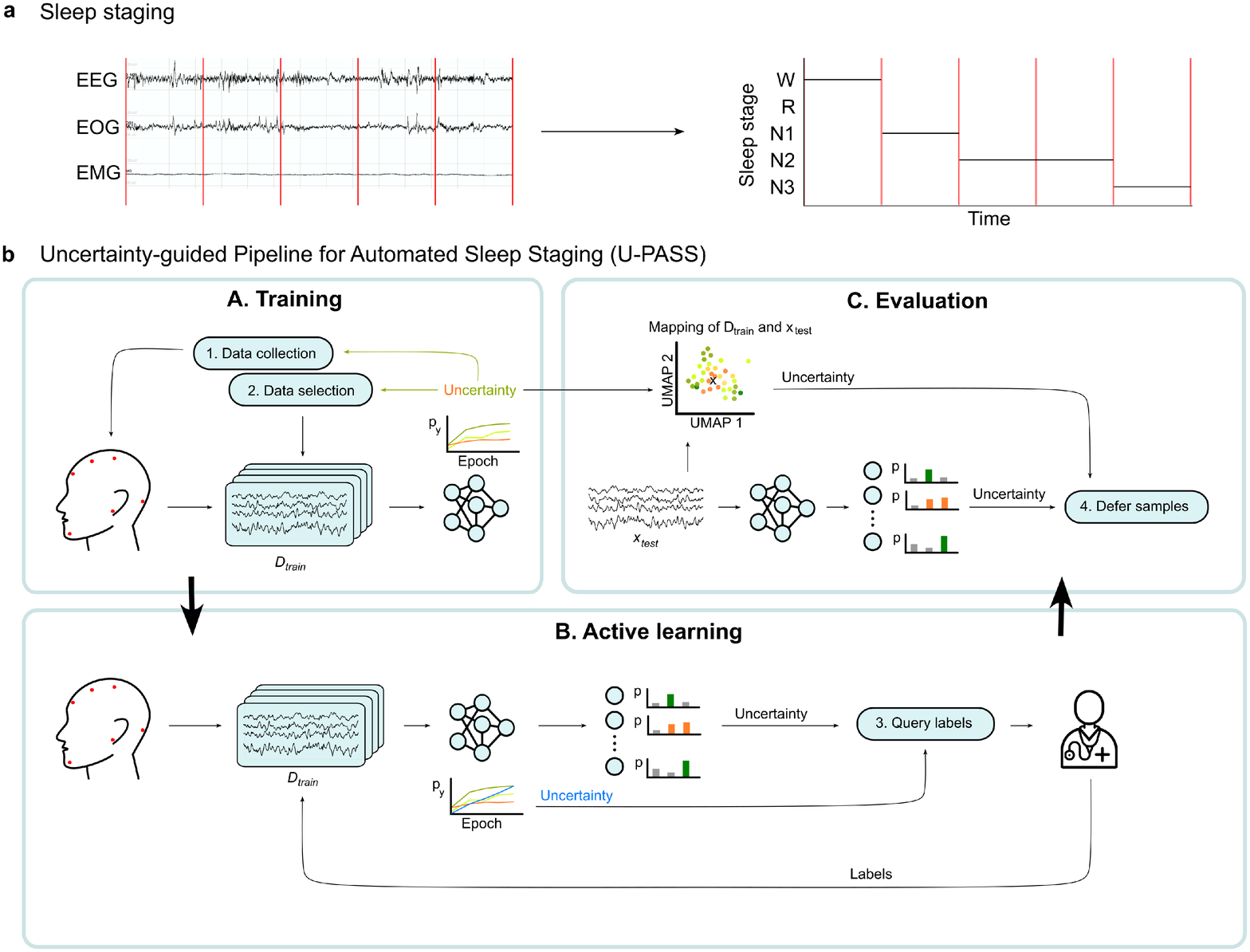}
\caption{(a) Illustration showing the classification problem of sleep staging. The recorded physiological signals are segmented in 30-second segments. Each segment is labeled as one of five sleep stages by SeqSleepNet \cite{Phan2018a}, a deep neural network architecture. (b) Full uncertainty-guided pipeline for sleep staging (U-PASS). The pipeline is more generally applicable to machine learning for clinical applications. \textbf{A.} At training time, data uncertainty is tracked to (1) select which channels (or features) to use, and (2) remove uninformative data. \textbf{B.} After training, active learning is used to finetune the model. Model uncertainty is tracked to (3) identify the most informative samples to query. 
\textbf{C.} At evaluation or deployment time, the most uncertain samples are deferred to the clinician (4). Used symbols: $p$ is the model output, $p_y$ is the model output for the true class $y$, $D_{train}$ is the training dataset and $x_{test}$ is a test sample.}
\vspace{-10pt}
\label{fig:pipeline}
\end{figure*}

The gold standard for diagnosing sleep-wake disturbances is by performing a polysomnography (PSG) study, which involves a fully monitored overnight stay in the hospital. Various physiological signals are recorded, including electroencephalography (EEG), electro-oculography (EOG) and electromyography (EMG) signals. These full-night recordings are segmented into 30-second segments, which are then labeled by trained clinicians. Each segment is labeled as one of five sleep stages according to clinical standards \cite{article, Kales1968a}. This so-called sleep staging process (see Figure \ref{fig:pipeline}a) is labor-intensive and subject to inter-rater variability, so automated approaches have been researched to help humans in this task. 
Numerous deep learning methods for sleep staging have recently been developed \cite{Phan2018a, Phan2020, Guillot2021a, Perslev2021}. 
These state-of-the-art sleep staging models achieve a similar performance to human experts. 
However, state-of-the-art models are mostly validated on healthy adults, and typically perform poorly on diseased subjects \cite{Phan2021}. Disease influences brain activity and, consequently, PSG recordings, creating uncertainty in the data. Moreover, models trained on data acquired with one measurement setup or device may not generalize well to data obtained using a different measurement protocol, resulting in the well-known distribution shift problem \cite{Eldele2023, Heremans2022b, Phan2019a, Pathak2021, Seo2020, Sun2017}. Hence, measurement factors can also contribute to uncertainty. Leveraging uncertainty estimation tools can help identify these uncertainties, which are likely to cause poor-quality predictions. 
As such, uncertainty estimates enhance model transparency and allow to actively query a sleep expert to label uncertain samples. Hence, by integrating uncertainty estimation into sleep staging, our uncertainty-guided deep learning pipeline for sleep staging (U-PASS) improves both trust and performance of the model.

\section{Methods}
\subsection{Data and study cohorts} \label{methods-data}
This study uses PSG recordings 
of 90 patients \cite{Heremans2022b}. It was recorded at the sleep laboratory of the University Hospitals Leuven (UZ Leuven) from January 2021 to September 2022. We included all patients over 60 years of age with suspicion of sleep apnea. This is a common sleep disorder in which breathing repeatedly stops during sleep. For each patient, the PSG recording includes EOG, EMG and a total of 6 EEG channels, all measured at 500 Hz. 45 recordings are used as a train set, and 45 recordings are used as the test set and for active learning. 
The dataset is manually annotated by an expert sleep scorer according to the AASM standard \cite{article}. 
The sleep study was conducted in accordance with the Declaration of Helsinki, and the protocol with registration number S64190/B3222020000148 was approved by the Ethics Committee Ethische Commissie Onderzoek UZ/KU Leuven. 

\subsection{Model and training}
U-PASS can be applied to any machine learning model and problem, as we designed it to be model agnostic, with no requirements or adaptations to the network architecture or training procedure. In this paper, we use it in a sleep staging context with a state-of-the-art sleep staging model, SeqSleepNet \cite{Phan2018a}. This deep neural network has a sequence-to-sequence classification scheme. It transforms a sequence of adjacent raw data segments into a corresponding sequence of outputs, in this case sleep stages. The inputs are presented as time-frequency images, spectrograms of each 30-second segment. The sequence length is a parameter, which is fixed to 10 in this study. SeqSleepNet is a hierarchical architecture, composed of a block of layers that processes individual segments, and a block that works at the sequence level. The segment processing block comprises a number of frequency filters, followed by an recurrent neural network (RNN) on the segment level. The outputs from the segment processing block are then concatenated into a sequence of inputs for the sequence processing block. A sequence-level RNN transforms this input sequence into an output sequence, which is mapped to a sequence of sleep stages by a softmax layer. The optimization and training parameters in this study are identical to those in the original paper \cite{Phan2018a}: L2 regularization is applied, and the Adam optimizer is used with a learning rate of $10^{-4}$. \\

\subsection{Formulation and visualization of uncertainty} \label{methods-formulation}

U-PASS utilizes insights from two recent methods that examine the characterization of uncertainty in training data. These methods analyze the training dynamics, which refers to the behavior of individual samples during the model training process. 
Data Maps \cite{Swayamdipta2020} maps instances by tracking predictive confidence and model uncertainty (epistemic uncertainty) during training. Data-IQ \cite{seedat2022dataiq} on the other hand maps instances using predictive confidence and data uncertainty (aleatoric uncertainty). Both methods use mappings to stratify the training dataset into subgroups of easy, hard and ambiguous data. We further refer to instances with high epistemic uncertainty as \emph{model-ambiguous} samples. 
These are examples for which the prediction variability of the model is highest during training. They are important to improve the performance and generalizability of a model \cite{Swayamdipta2020}. 
On the other hand, instances with high aleatoric uncertainty are further referred to as \emph{data-ambiguous} samples. 
These are samples for which predictions are the most uncertain on average, and on which the model systematically underperforms \cite{seedat2022dataiq}.\\


Let us formally define the characterization of model-ambiguous samples and data-ambiguous samples. Consider a typical supervised learning setting for a classification problem, where inputs $x \in \mathbb{R}^d$ need to be assigned to classes $y \in \mathbb{N}$. The goal is to learn a model $f_\theta$ which maps inputs to outputs by assigning a probability to each class given the input: $f_\theta(x)=P(Y|X=x,\Theta=\theta)$. Let $\theta$ designate a certain instantiation of the model parameters. The model is iteratively trained for $e^*$ training epochs, resulting in $e^*$ different instantiations of the model parameters $\{ \theta_1, \theta_2, ... ,\theta_{e^*}\}$. Let $\Theta \sim P_{emp} (\{ \theta_1, \theta_2, ... ,\theta_{e^*}\})$ be a random variable with an empirical distribution over the model parameters throughout the training process. 
$\mathcal{P}(x,\theta) = [f_\theta(x)]_y$
is the model's output probability for the true class label. Using the variance as an uncertainty metric, we can decompose the uncertainty by the law of total variance. The epistemic uncertainty of the prediction is ${v}_{ep}=\mathbb{V}_{\Theta} [ \mathcal{P}(x,\Theta)]$. The variance is evaluated over the model outputs during training, hence epistemic uncertainty captures the variability of model outputs, representing model uncertainty. The aleatoric uncertainty is ${v}_{al}=\mathbb{E}_{\Theta} [ \mathcal{P}(x,\Theta) (1-\mathcal{P}(x,\Theta)) ]$ \cite{seedat2022dataiq}. In the aleatoric uncertainty, the variance is computed over the prediction. Hence, this represents the data uncertainty. Using the empirical distribution of $\Theta$, the uncertainties are calculated as follows \cite{seedat2022dataiq}:
\begin{equation}
{v}_{ep}= \frac{1}{e^*} \sum_{e=1}^{e^*} ( \mathcal{P}(x,\theta_e) -  \mathcal{\overline{P}}(x) )^2 ,
\end{equation}
\begin{equation}
{v}_{al}= \frac{1}{e^*} \sum_{e=1}^{e^*}  \mathcal{P}(x,\theta_e) (1- \mathcal{P}(x,\theta_e)),
\end{equation}

with $\mathcal{\overline{P}}(x)=1/e^* \sum_{e=1}^{e^*} \mathcal{P}(x, \theta_e)$. \\

We define model-ambiguous data as data with a high ${v}_{ep}$, and data-ambiguous samples as data with high $v_{al}$. In making these two groups, we use the definitions of `ambiguity' proposed in Data Maps \cite{Swayamdipta2020} and Data-IQ \cite{seedat2022dataiq}, respectively. We can further stratify the training dataset using the concept of predictive confidence as the mean output probability for the ground-truth-class ${c}=\mathcal{\overline{P}}(x)$, i.e. the confidence in the correct class. This allows us to define easy-to-classify samples as instances with high confidence and low ambiguity and hard-to-classify samples as instances with low confidence and low ambiguity. \\

\subsection{U-PASS pipeline}
Based on the uncertainty quantification methods from Section \ref{methods-formulation}, we build a full uncertainty-guided pipeline, outlined in Figure \ref{fig:pipeline}b. \textbf{A.} First, we tailor and improve the training dataset. Even before collecting a full dataset, we can start training a model using various measured signals and monitor the data uncertainty to pinpoint which signals are effective in decreasing uncertainty. This information can aid in deciding which modalities and channels should be recorded. There is a trade-off between the cost of data acquisition and the amount of uncertainty that can be tolerated to achieve a clinically useful result. After choosing a measurement setup, a full dataset is collected. We can then train a model on this dataset, and track the data uncertainty to identify the most data-ambiguous samples. These samples are removed from the training dataset, as they confuse the model instead of improving its performance. \textbf{B.} After training, the second step in U-PASS is to finetune the model using active learning. 
The model uncertainty can be used to identify for which recordings finetuning is the most useful, i.e. which recordings have the highest learning potential. Indeed, model uncertainty measures how much predictions of the model vary during training. 
In this work, we achieve finetuning with a semi-supervised transfer learning technique, using the additional labels supplied through active learning. 
\textbf{C.} Lastly, we can use uncertainty at evaluation or test time. A simple uncertainty measure based on uncertainty of the neighboring training samples can be used to identify uncertain test samples. Instead of making an incorrect prediction on these samples, it is better to defer them to a clinician. The three principled uncertainty-guided steps in U-PASS result in a reliable and robust machine learning pipeline that is tailored for clinical environments. This framework enables effective interaction between clinicians and the model, thus enhancing clinical applicability. We delve into each next. 

\subsubsection{Data collection and data selection} \label{methods-select}

In the data collection experiments, we train the model on different numbers of channels to identify the best measurement setup. In the data selection experiments, we train on the best channels based on the data collection experiment, removing different amounts of ambiguous training data in each experiment. In both the data collection and data selection experiments, the model is trained on the full training dataset for 10 training epochs. Early stopping is applied using a validation dataset consisting of 2 recordings, retaining the model with the best performance on the validation set. As the training dataset consists of 45 recordings, each experiment is repeated $\lfloor45/2 \rfloor =23$ times, with different recordings in the validation set for every repeat. Accuracies are computed on the separate test set consisting of 45 recordings. The test set accuracies and training set uncertainty measures reported in Figure \ref{fig:case1}a  are computed as averages over these 23 repeats.


\subsubsection{Active learning} \label{methods-al}
The active learning experiments start from the model obtained with the optimal setting based on the channel and data selection experiments. Using active learning to query informative labels, the model is then finetuned to personalize it to individual patients with the semi-supervised adversarial domain adaptation approach from \cite{Heremans2022b}. The training set of 45 patients is used as the source dataset, and each selected individual recording of the test set as the target dataset for adversarial domain adaptation. This experiment is performed 23 times, with each of the trained models (see \ref{methods-select}). 
The labels for the semi-supervised training approach are acquired through the following active learning scheme. Before every training epoch, 1\% of samples (roughly 10) are queried. The labels of these samples are used during that training epoch, along with the labels cumulatively acquired in previous epochs. The basic unsupervised adversarial domain adaptation is thus augmented by adding a supervised loss from the queried labels. The final model obtained after 10 training epochs with querying is retained. The instantiated querying strategy is a simple uncertainty-based querying strategy based on the prediction entropy of the current model at every epoch $H [{P}(y|x,\Theta_{f})]$. Any other querying strategy could also fit in the U-PASS pipeline. \\

The model uncertainty is a measure for how much the model performance varies while training on certain data, and can hence be seen as a measure for learning potential. Therefore, we focus the labeling efforts of the clinician on the recordings with the highest model uncertainty. 
Section \ref{methods-formulation} describes how to estimate model uncertainty using the variance of the prediction for the correct class as an uncertainty metric, specifically for training data that has ground truth labels available. 
In our active learning scenario, labels are not available. However, we can still analyze training dynamics to estimate uncertainty, using entropy as an alternative uncertainty metric \cite{olmin2022uncertainty}. The definition of epistemic and aleatoric uncertainty with this uncertainty metric are the following:
\begin{equation}
{v}_{al}=\mathbb{E}_{\Theta} [ H [{P}(y|x,\Theta)] ]
\label{v_al}
\end{equation} 
\begin{equation} 
{v}_{ep}=H[\mathbb{E}_{\Theta} [ {P}(y|x,\Theta) ]] - \mathbb{E}_{\Theta} [ H [{P}(y|x,\Theta)] ]
\label{v_epi}
\end{equation}

with $H$ the entropy metric, and $P(y| x,\Theta)$ the output distribution. We train the model on every recording using unsupervised adversarial domain adaptation, estimate the model uncertainty $v_{ep}$ based on the training dynamics in this unsupervised training step, and select the recordings with the highest $v_{ep}$ for active learning.

\subsubsection{Deferring ambiguous test samples} \label{methods-defer}
In the evaluation experiments, we used the personalized models obtained in the AL experiments with model uncertainty as the query strategy. As uncertainty metrics based on training dynamics cannot be used at deployment time, we must rely on post-hoc uncertainty metrics. Several post-hoc uncertainty metrics were compared, including two that were based on the output of the model: the entropy of the output distribution and the maximum output probability (i.e. the output probability of the predicted class). Two other metrics were distance-based. One metric measured the distance of the test sample to its $n$ closest train samples, while the other used the distance of the test sample to the $n$ closest training samples for each separate class, and was calculated as the proportion of the smallest distance to the second-smallest distance. The remaining three metrics calculated uncertainty by averaging metrics of the $n$ closest training samples, weighted by their distance to those training samples. One metric used the average confidence of the closest training samples, the second used the average data uncertainty, and the third used the average model uncertainty. \\

Every uncertainty metric for evaluating uncertainty on the test set has its own scale, and it is up to the domain experts to define what level of uncertainty and how many mistakes can be tolerated. Therefore, we designed this experiment by ranking test set samples from most to least uncertain, and plotting the accuracy corresponding with the z\% most uncertain test samples. The uncertainty measures were then compared by averaging these accuracies over the 23 trained models. 

\section{Results}


\subsection{Insights from uncertainty}
\begin{figure*}[]
    \centering
    \includegraphics[width=\textwidth]{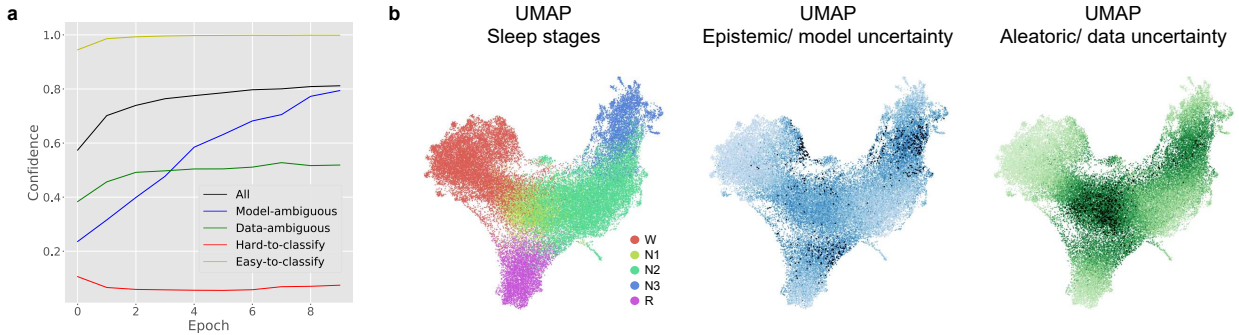}
    \caption{
    (a) The average training dynamics for the 1\% most data-ambiguous, model-ambiguous, easy-to-classify and hard-to-classify samples. (b) UMAP visualization of the feature distribution resulting from training a sleep staging network. The left plot shows the sleep stages in this distribution. In the middle plot, the tint indicates the model uncertainty, and the 1\% most model-ambiguous are shown in black. In the right plot, the tint indicates the data uncertainty, and the 1\% most data-ambiguous are shown in black. 
    }
    \label{fig:umaptrdy}
\end{figure*}

First, we illustrate how the two types of uncertainty are manifested in our data in Figure \ref{fig:umaptrdy}. Figure \ref{fig:umaptrdy}a shows the training dynamics for the 1\% most model-ambiguous and data-ambiguous samples, as well as for the 1\% most easy-to-classify and hard-to-classify samples. Model-ambiguous samples show the greatest learning curve during training, while data-ambiguous samples are characterized by a flat learning curve. The distributions of the two types of ambiguity are visualized in Figure \ref{fig:umaptrdy}b. The figure shows two-dimensional UMAP embeddings of the training data, with the different colors indicating the five sleep stages in the left figure, and heatmaps showing the model and data uncertainty in the middle and right figure, respectively. In both heatmaps, the 1\% most ambiguous samples are highlighted in black. 
The 1\% most model-ambiguous samples 
are found mostly at the borders and outside the main training data distribution, as well as on class boundaries. Indeed, atypical and out-of-distribution samples are more challenging for the model at first, but the model can learn these samples through more training. The most data-ambiguous samples are centered in sleep stage N1, which is the least well defined and hardest to classify. Data uncertainty is also higher on the class boundaries, where some samples contain characteristics of multiple sleep stages. \\ 

\subsection{Data collection}

\begin{figure*}
\begin{subfigure}[b]{\textwidth}
    \centering
    \includegraphics[width=\textwidth]{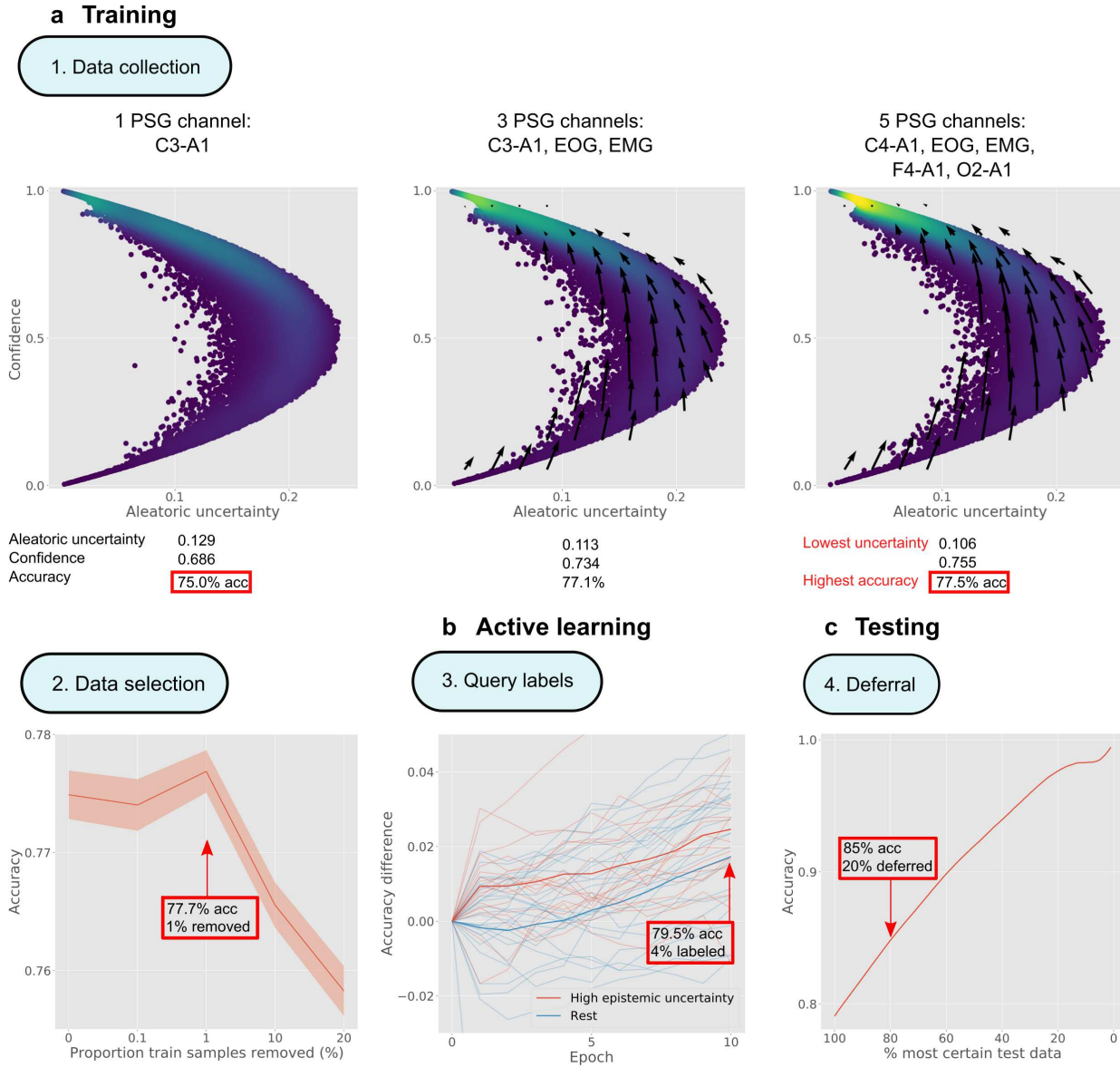}
\end{subfigure}
\caption{U-PASS applied to hospital-based polysomnography (PSG) data. The resulting accuracy after every consecutive step is indicated with a red box. (a) During training, the training dataset is tailored using data collection and data selection based on data uncertainty. In the data collection step, samples move to the upper-left corner with high confidence and low data uncertainty when more channels are used. We thus proceed with all five channels. In the data selection step, the optimal test result is achieved when removing the one percent (roughly 440) most data-ambiguous training samples. (b) Active learning is used to adapt the model to individual recordings. 1\% of samples is queried in each of the 10 training epochs. We only perform active learning on recordings characterized by high model uncertainty. On the rest of the recordings, it doesn't show a large improvement. (c) At deployment time, uncertainty-based deferral increases the test accuracy to the desired level of 85\%. Samples with uncertainty values below the threshold are referred to a clinician for manual labeling.}
\label{fig:case1}
\end{figure*}

U-PASS starts with evaluating the data uncertainty during data collection to assess which signals should be acquired and used. 
Figure \ref{fig:case1}a shows the confidence and data uncertainty in the PSG training datasets, mapped by using Data-IQ \cite{seedat2022dataiq}. Each 30-second sample is one dot, and the density of samples is represented by the brightness of colors. The figure demonstrates that most samples are in the upper left corner, characterized by high confidence and low data uncertainty (easy-to-classify). Some samples have high data uncertainty (data-ambiguous), and a minority of samples is characterized by low data uncertainty and low confidence (hard-to-classify). The concentration of samples in the data-ambiguous region decreases when going from one channel to three channels, and decreases again from three channels to five channels. This is also evident from the average data uncertainty over the whole training dataset, which is shown under the plots. The average confidence increases accordingly, as samples go from being data-ambiguous to easy-to-classify. As a result, the accuracy on the test set also improves, which is the final proof that reducing the data uncertainty is helping the outcome. \\

We conclude that tracking the data uncertainty allows to compare the quality of different measurement setups for sleep monitoring. Indeed, a decrease in data uncertainty with the addition of a channel indicates that this channel adds valuable information. It can advise users in selecting the optimal setup that will result in the best accuracy at deployment time. 


\subsection{Data selection} 
We can curate our training dataset, by discarding from it the most data-ambiguous training samples. Figure \ref{fig:case1}a shows how removing these ambiguous data affects the model's test performance on an independent test set. There is a trade-off to be made between removing samples that confuse the model and removing too many samples so that useful information is discarded from the training dataset. In this case, the optimal point is reached when discarding 1\% of training samples. \\ 

\begin{figure*}
\centering
\begin{subfigure}[b][][t]{0.8\textwidth}
     \centering
\includegraphics[width=\textwidth]{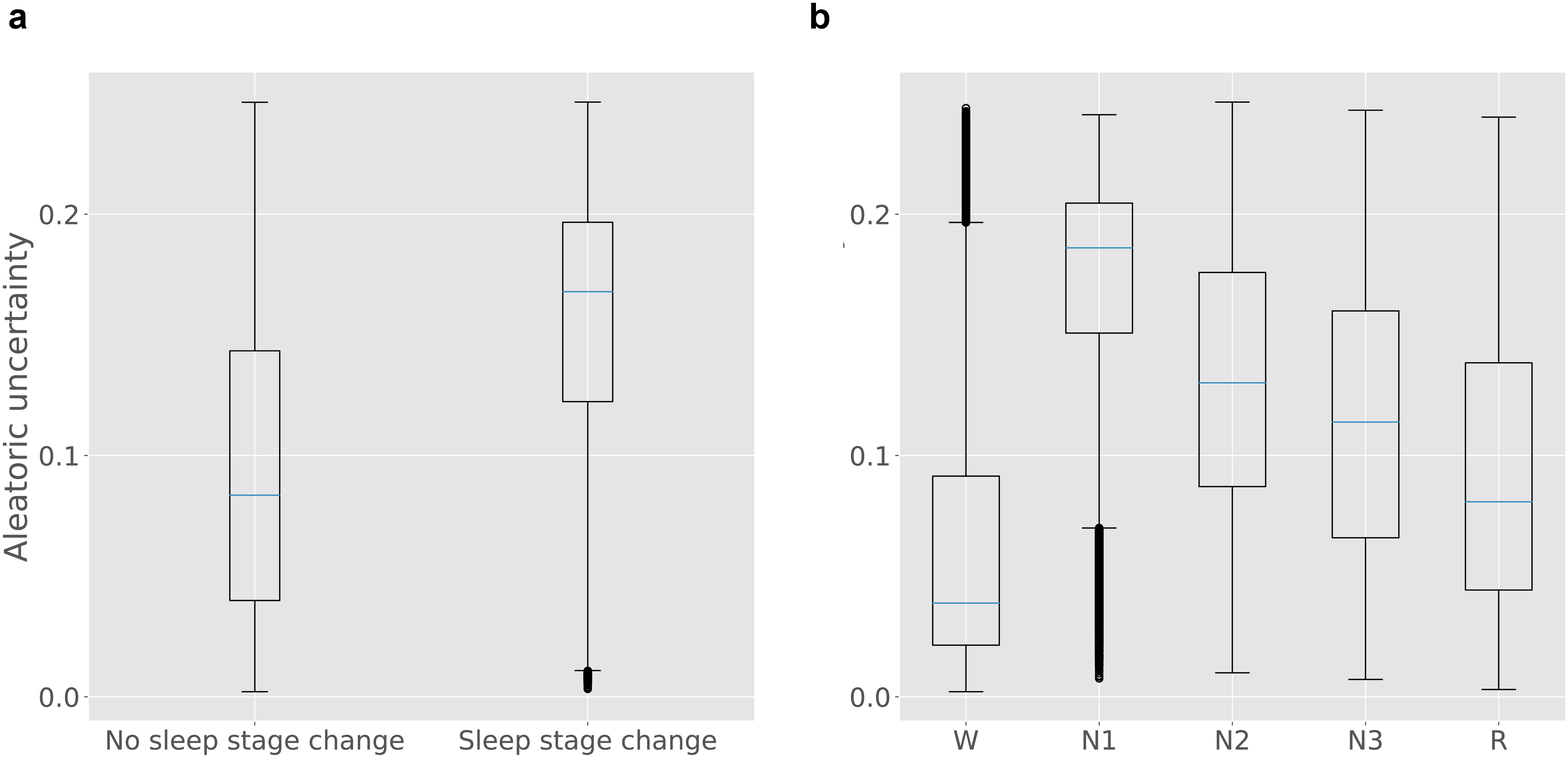}
\end{subfigure}
\caption{Parameters causing data uncertainty in the PSG training dataset. (a) The data uncertainty per segment is higher for segments on sleep stage transitions than for segments that are not on sleep stage transitions. (b) The data uncertainty per segment, aggregated per sleep stage, shows that wake segments have low data uncertainty and N1 segments have the highest data uncertainty.}
\label{fig:expl_tr}
\end{figure*}

Data uncertainty in sleep staging can arise from biological sources (e.g. age, pathology) and  sources related to the measurement technique (e.g. segmentation, interference) \cite{VanGorp2022}. 
Figure \ref{fig:expl_tr}a shows that data uncertainty per segment is significantly higher for segments on sleep stage transitions than for segments that are not on sleep stage transitions. This source of uncertainty is hence related to the discretization caused by segmentation and the continuous nature of sleep, with features from multiple sleep stages in a single segment. 
Figure \ref{fig:expl_tr}b shows that wake segments have the lowest data uncertainty, and N1 segments have the highest data uncertainty. The fact that different stages of sleep are characterized by different levels of uncertainty is related to how clearly sleep stages are recognized and defined, which is influenced by sleep scoring rules and the biology of sleep. We find evidence for this in the fact that the interrater agreement between human scorers shows the same trends as are observed through data uncertainty \cite{Lee2022}. 

%
%

\subsection{Active learning}
\begin{figure*}[]
\centering
\begin{subfigure}[b]{0.9\textwidth}
    \centering
    \includegraphics[width=\textwidth]{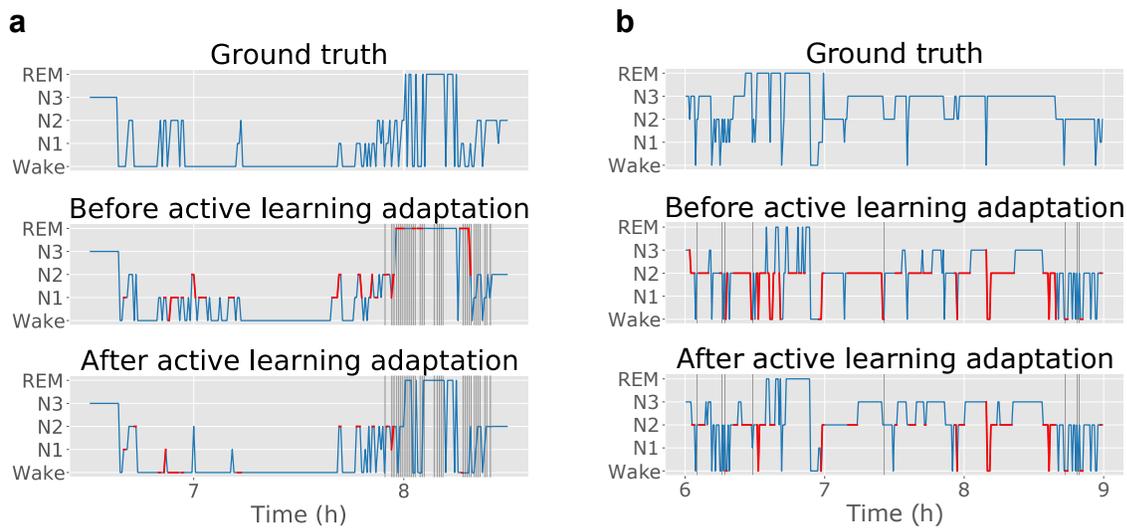}
\end{subfigure}
\caption{Hypnograms (sleep stages over time) for snippets of two different recordings, demonstrating how personalization through active learning (AL) improves the performance. Queried samples are shown in grey, errors are shown in red. (a) Before AL, there are a lot of incorrect N1-sleep classifications around hour 7, and the jumps between REM-sleep and Wake around hour 8 are missed. After AL, most of these mistakes are corrected. Although the queried samples are centered around hour 8, the mistakes in N1-sleep around hour 7 are partly corrected too. (b) Both the episode of REM-sleep between hour 6.5 and 7, and the episodes of N3-sleep between hour 7 and 8.5 are more accurately predicted after AL. Overall, we see that both the queried samples and the unlabeled parts of the recording are better classified after AL in both examples. }
\label{fig:hypno_psg_ALimprovt}
\end{figure*}

%
Once the model is trained on the improved
training set, U-PASS evaluates model uncertainty to assess on which recordings the model should be finetuned through active learning. We use active learning to personalize sleep staging models to those individual patients. 
Figure \ref{fig:case1}b shows the average improvement from active learning to individual PSG recordings. The improvement is shown for the selected 40\% recordings with the highest model uncertainty and for the rest of the recordings. Model uncertainty clearly points to the recordings that benefit the most from the active learning adaptation. Figure \ref{fig:hypno_psg_ALimprovt} shows two examples of how active learning improves the predictions in both labeled and unlabeled samples.\\

\subsection{Deferring uncertain test samples}

Lastly, the finetuned model is deployed on unseen data. U-PASS then uses uncertainty measures to defer test samples to an expert when predictions are uncertain.
Figure \ref{fig:case1}c 
shows the accuracy on the 100\% to 1\% most certain test samples. This is equivalent to the accuracy when removing 0\% to 99\% of the most uncertain test samples. The higher this accuracy, the better the uncertainty estimation, as the uncertainty metric should retain the most accurate predictions and defer the least accurate ones. Different post-hoc uncertainty metrics were applied.  Figure \ref{fig:case1}c only shows the best-performing one: the weighted average confidence of the $n$ closest training samples. 
The performances of all the uncertainty metrics are shown in Supplementary Figure \ref{fig:reject_extra}. \\

The results in Figure \ref{fig:reject_extra} show that different uncertainty metrics perform best depending on the threshold set on the tolerated uncertainty. 
If we require 85\% accuracy, the confidence of neighboring training samples achieves the best performance, with 20\% 
of samples deferred. For reference, 85\% accuracy is a great performance, as the inter-rater agreement in sleep staging is estimated to be 82.6\% \cite{Rosenberg2013, Lee2022}. When we exclude samples that lie on sleep stage transitions, the accuracy improves from 85\% to 89\%.
\\



\begin{figure*}[]
\centering
\begin{subfigure}[b]{0.6\textwidth}
    \centering
\includegraphics[width=\textwidth]{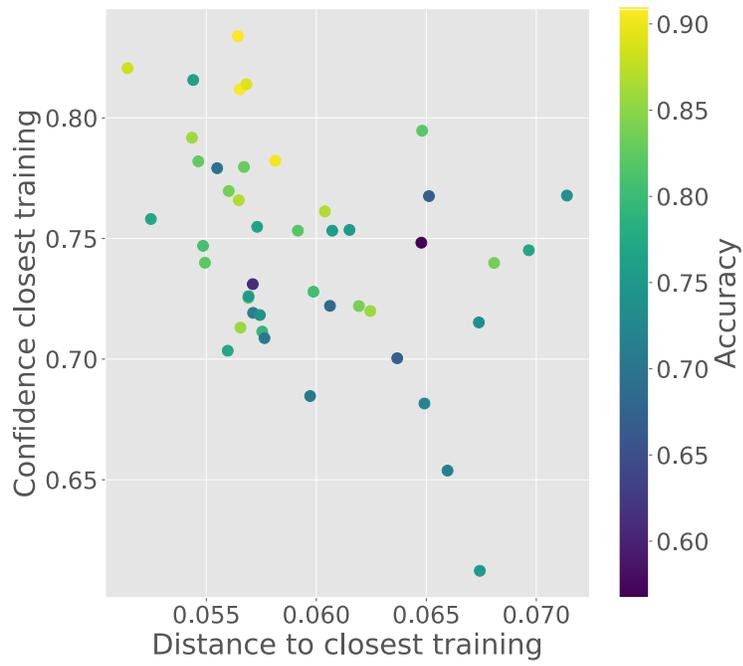}
\end{subfigure}
\caption{
The accuracy on the test data depends on both the distance from the training data and the uncertainty of those training data. For every patient, this figure shows the accuracy, confidence of the closest training samples and the distance to those training samples. These metrics are averaged over all the samples of the patient's recording, so every patient is represented by one dot.  
}
\label{fig:expl_rej1}
\end{figure*}

As clear from Figure \ref{fig:reject_extra}, the difference between the entropy metric and metrics based on the uncertainty in the training set is mostly quite small, and depends on the uncertainty threshold we set. However, the metrics based on uncertainty in the training set have an important advantage compared to the decision entropy, as they provide a reason for deferral. Indeed, the distance-weighted confidence of neighboring training samples is a product of the distance from the training samples and the uncertainty of close training samples. Figure \ref{fig:expl_rej1} visualizes both factors. The Spearman correlation coefficient shows that accuracy is negatively correlated with distance from the closest training samples (r(43)=-.40, p=.0067) and positively correlated with the confidence of the closest training samples (r(43)=.51, p=.00032). The two factors detect, respectively, out-of-distribution samples, and samples lying close to uncertain regions, e.g. decision boundaries.\\

The effect of applying U-PASS for sleep staging is summarized in the red boxes of Figure \ref{fig:case1}, which show the resulting accuracy after each step in the pipeline. This clearly demonstrates how U-PASS leverages uncertainty in training, active learning and deployment to improve the accuracy at every stage. 

\section{Discussion}
We have developed and validated U-PASS, an uncertainty-guided pipeline for automated sleep staging. Although we applied U-PASS to sleep staging, the pipeline is generally applicable to any machine learning problem. It is particularly useful for machine learning in healthcare, which has strong requirements for reliability, trust, and benefits from user interaction. By utilizing data uncertainty and model uncertainty at different stages in the data processing pipeline (training, active learning and deployment), we have shown how U-PASS improves predictions step by step. It can easily be tailored to the accuracy needs of any particular application, by adapting the amount of data queried and deferred in the active learning and deployment phase, respectively. \\

We have validated our pipeline in PSG data of a real patient population of elderly subjects with suspicion of sleep apnea. 
Many performant deep learning models have been developed for sleep staging on PSG, but they are developed as static, stand-alone input-output machines. This limits their performance and usability in the clinic. Our contribution consists of building a pipeline in which such deep learning models can be integrated, boosting their performance and allowing to choose the desired performance level.  \\

The performance gains achieved by applying U-PASS to a state-of-the-art sleep staging model learning are shown in Figure \ref{fig:case1}. The first two parts of the pipeline, data collection and data selection, are focused on tailoring the training dataset to achieve optimal results. Both steps decrease the data uncertainty in different ways. In the data collection step, data uncertainty is estimated to find out which features (or channels in the case of sleep staging) need to be collected. Once the data acquisition protocol is fixed and the training set is acquired, the data selection step filters out the data with the highest data uncertainty, as these can deteriorate model performance. In our experiments, the data collection step shows the largest improvement by going from one EEG channel to five channels. The data selection step only increases the outcome by a relative 0.4\%, a small but significant percentage. This maximum improvement is attained when 1\% of the data are discarded. Our training dataset is not overly large, so we hypothesize that the optimal percentage of data to discard and the corresponding improvement may change depending on the dataset size. In future work, we could investigate the influence of the type, `cleanness' and size of training data on the data selection. \\ 

The last step of U-PASS, deferring ambiguous samples, allows us to choose a tolerated level of uncertainty, or a desired accuracy for a specific task. This comes at a labeling cost, similar to the active learning step of the U-PASS pipeline. 
The difference between the two steps is the type of uncertainty they tackle. The only uncertainty that can be resolved through active learning is model uncertainty: uncertainty coming from a lack of knowledge of the model. 
Data uncertainty, on the other hand, is inherent to the data and can hence not be resolved through learning. The only possible mechanism to cope with high data uncertainty at deployment time is deferral to an expert. In theory, doctors do not know more than a machine learning model in cases of pure data uncertainty. 
However, they can take the right course of action, whether that be ignoring the data, performing a new measurement, or fixing a broken electrode. We conclude that both active learning and deferring help to increase the accuracy to the desired level at a certain labeling cost. Since they tackle different types of uncertainty, they should be combined for optimal performance and usability.\\ 
%

Along with performance gains, the uncertainty estimation methods in the U-PASS pipeline also provide some interesting clinical insights into the sleep data. Figure \ref{fig:umaptrdy}b and Figure \ref{fig:expl_tr} shows that the data uncertainty in segments on sleep stage transitions is higher than for others, and that sleep stage N1 is characterized by more uncertainty than other sleep stages. This data uncertainty is by definition inherent to the data and can not be resolved by more training. These results are also consistent with the agreement between human sleep scoring experts \cite{Lee2022}. As such, the objectivity of uncertainty metrics can guide clinical practitioners and experts developing sleep scoring guidelines \cite{article} to define better rules. For example, changing the 30-second segmentation to a more fine-grained segmentation should benefit the data uncertainty. Furthermore, the less ambiguous the sleep stages (through clearer staging guidelines or even by changing the sleep stages themselves), the less data uncertainty. An additional insight we have gained from plotting the sleep features (Figure \ref{fig:umaptrdy}b) is that sleep stages don't seem to be discretely separated, but rather lie on a continuous spectrum. Recent works have advocated modelling sleep as a continuous process \cite{Younes2015, Cesari2021, Hermans2022}, which may be a more biologically accurate representation. Hence, uncertainty estimation methods can inform the medical field on how to best describe and define sleep.\\

 In conclusion, U-PASS is a machine learning pipeline that integrates uncertainty-informed curation of the training set, uncertainty-based active learning to incorporate a clinician’s feedback and deferral of uncertain decisions to a clinician.
 As such, all the facets of uncertainty and benefits of uncertainty estimation are harmonized in a single framework. This optimizes the machine learning pipeline and unlocks its potential in a clinical setting by adding safety guardrails to the process. 
 Furthermore, U-PASS has the potential to provide medical practitioners with valuable insights into their data, offering a deeper understanding of sleep biology and potentially other clinical applications. Overall, U-PASS represents a promising approach to enhancing the reliability and safety of machine learning systems in critical fields such as healthcare.

\section{Supplementary material}

\begin{figure*}[h]
\begin{subfigure}[][][t]{\textwidth}
    \centering
    \includegraphics[width=\textwidth]{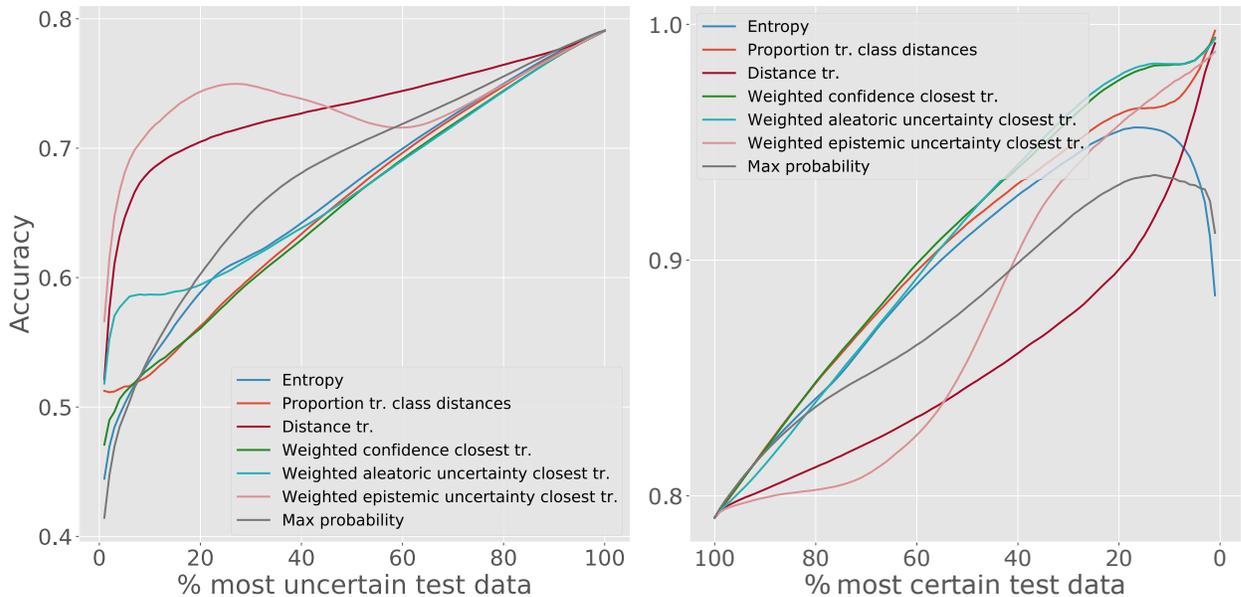}
\end{subfigure}
\caption{Extended results showing the performance of the different measures of uncertainty for deferring uncerrtain test samples. Left is the accuracy for the x\% most uncertain samples. The better the uncertainty estimate, the lower this figure. Right is the accuracy for the x\% most certain samples. The better the uncertainty estimate, the higher this figure. 
Depending on the threshold and scenario, different uncertainty measures perform best. } 
\label{fig:reject_extra}
\end{figure*} 


\bibliographystyle{IEEEtran.bst}
\bibliography{refs_mendeley}

\end{document}